\pdfoutput=1

\documentclass[conference,a4paper]{APSIPA2021}
\usepackage[numbers,sort&compress]{natbib}
\usepackage[colorlinks,citecolor=black]{hyperref}
\usepackage{multirow}
\usepackage{amsmath}
\usepackage[psamsfonts]{amssymb}
\usepackage{amsfonts}
 \usepackage{booktabs}
\usepackage{amsxtra}
\usepackage{threeparttable}
\usepackage{graphicx}
\usepackage{float}
\usepackage{bm}
\usepackage{mlmath}
\usepackage{geometry}
\usepackage[normalem]{ulem}
\useunder{\uline}{\ul}{}
\usepackage{makecell}

\usepackage{pifont}

\usepackage{bbding}

\begin{document}

\title{\texorpdfstring{Decoupled Latent Flow Matching for Few-Step Joint Vocal-Accompaniment Separation}{Decoupled Latent Flow Matching for Few-Step Joint Vocal-Accompaniment Separation}}

\author{
    Lishi ZUO, Youzhi TU, Lu YI, Zezhong JIN, Chongxin GAN, Man-Wai MAK, KongAik LEE\\
\texttt{Dept. of Electrical and Electronic Engineering,}\\
\texttt{ Hong Kong Polytechnic University, Hong Kong SAR, China}\\
\texttt{E-mail: lishi.zuo@connect.polyu.hk, youztu@polyu.edu.hk, luicerain@gmail.com,}\\
\texttt{zezhong.jin@connect.polyu.hk, chong-xin.gan@connect.polyu.hk,}\\
\texttt{man.wai.mak@polyu.edu.hk, kong-aik.lee@polyu.edu.hk}
}

\maketitle
\thispagestyle{empty}
\begin{abstract}
Generative modeling provides a flexible way to model mixture-conditioned source distributions, but iterative diffusion and flow matching models are costly for long music signals. 
This paper studies joint vocal-accompaniment separation through latent flow matching, where a pretrained variational autoencoder (VAE) maps mixtures and sources into a compact latent space and a flow matching model generates vocal and accompaniment latents jointly. 
The proposed framework decouples semantic separation from acoustic velocity prediction through a Separation Encoder and a Velocity Decoder. 
To reduce sampling cost, we further apply latent adversarial post-training inspired by Flow2GAN for few-step generation. 
Experiments show that latent adversarial refinement can improve perceptual and separation metrics under a reduced sampling budget.
\end{abstract}

\section{Introduction}

Music source separation is a fundamental task for music editing, remixing, and content creation \cite{musdb18,stoter2018sisec}. Vocal-accompaniment separation is especially challenging because the two target sources often share temporal, rhythmic, harmonic, and structural patterns within the same song. Modeling them separately can overlook these cross-source dependencies and hide quality trade-offs between the output streams. This motivates a joint formulation in which vocal and accompaniment are treated as correlated sources rather than unrelated prediction targets.

We study this problem from a generative perspective, not as a replacement for strong discriminative separators, but as a way to model mixture-conditioned source distributions. Flow matching provides a continuous-time generative framework for learning transformations from simple noise distributions to complex audio distributions \cite{lipman2023flowmatching,liu2023rectifiedflow}. The objective of this study is to examine whether joint latent flow matching can provide an efficient generative framework for vocal-accompaniment separation, and to analyze how semantic-acoustic decoupling and latent adversarial refinement affect source fidelity, source balance, and sampling efficiency.

Directly applying generative models to waveform-level music separation, however, is costly because long audio sequences require high-dimensional generation and repeated sampling. A latent formulation can reduce this burden by implementing the generative process in a compact learned representation \cite{rombach2022latentdiffusion}. In this work, a pretrained variational autoencoder (VAE) maps the mixture and target sources into latent representations, and the flow matching model jointly generates the vocal and accompaniment latents before decoding them back to waveforms.

Joint source modeling does not mean that all information inside the generative model should be optimized in the same way. Source-specific information separation and acoustic generation play different roles: the former determines what source information should be preserved from the mixture, while the latter determines how the corresponding acoustic states evolve during generation. This motivates us to investigate whether separating these roles can reduce optimization interference in latent flow matching. We therefore introduce a semantic-acoustic decoupling strategy that separates the Separation Encoder from the Velocity Decoder while keeping vocal and accompaniment jointly modeled in the same latent flow.

Although latent flow matching reduces the modeling complexity, inference still requires iterative integration of the learned velocity field. This sampling cost is undesirable for music separation, where long audio segments are common. To enable few-step generation, we further extend Flow2GAN-style adversarial acceleration \cite{yao2025flow2gan} to the latent space. The latent discriminator encourages few-step outputs to match the structure of real source latents, while the latent reconstruction objective preserves correspondence to the target sources. Because vocal and accompaniment are generated jointly, few-step acceleration must also avoid improving one source at the expense of the other. The proposed latent refinement is therefore designed to improve efficiency and to study how adversarial few-step training affects the balance between the two output streams.

Overall, the contributions of this work are summarized as follows:
\begin{itemize}
    \item \textbf{Joint latent generative modeling.}
    We formulate vocal-accompaniment separation as joint latent flow matching, allowing the two correlated sources to be generated simultaneously within a shared VAE latent space.

    \item \textbf{Semantic-acoustic decoupling.}
    We decouple source-specific representation extraction from acoustic velocity prediction using a Separation Encoder and a Velocity Decoder.

    \item \textbf{Latent few-step generation.}
    We introduce Flow2GAN-style adversarial post-training in the latent space for few-step source generation and show its effect on the quality trade-off between vocal and accompaniment.
\end{itemize}

\section{Related Work}
\subsection{Music Source Separation}
Music source separation has traditionally been dominated by discriminative models that directly estimate target sources from a mixture. 
Representative methods include spectrogram-domain models such as Open-Unmix \cite{stoter2019openunmix}, time-domain convolutional approaches such as Conv-TasNet \cite{luo2019convtasnet}, and hybrid or waveform-domain music separation method such as Demucs \cite{defossez2021demucs}. 
These methods have achieved strong separation performance, 
but they are usually optimized as direct predictors rather than explicit generative models of the target source distribution. 
In contrast, this work focuses on a generative formulation for vocal-accompaniment separation, with the goal of modeling the two correlated sources within a shared latent generative process rather than competing with large discriminative systems on separation metrics.

\subsection{Generative Modeling and Few-Step Acceleration}
Diffusion and score-based models have shown strong ability to model complex distributions \cite{ho2020ddpm,song2021score}, and flow matching provides a continuous-time generative framework based on learning velocity fields \cite{lipman2023flowmatching,liu2023rectifiedflow}. For high-dimensional audio generation, latent generative modeling can reduce the cost of generation by moving the process into a compact latent space \cite{rombach2022latentdiffusion}. 
Because standard diffusion and flow matching models still require iterative sampling, 
recent few-step methods reduce inference cost through consistency or adversarial distillation \cite{song2023consistency,sauer2024add}, 
average-velocity formulations such as MeanFlow and Improved MeanFlow \cite{geng2025meanflow,geng2026improvedmeanflow}, and distribution-level distillation such as Distribution Matching Distillation (DMD) \cite{yin2024dmd}. 
For audio generation, Flow2GAN combines flow matching with adversarial refinement to improve waveform quality \cite{yao2025flow2gan}. These directions motivate efficient generative separators, but leave open how latent few-step refinement behaves when vocal and accompaniment are generated jointly. Our work is most closely related to Flow2GAN, which focuses on GAN training in the waveform domain, whereas our method operates in the latent space.

\section{Methodology}
\begin{figure*}[t]
    \centering
    \includegraphics[width=\linewidth]{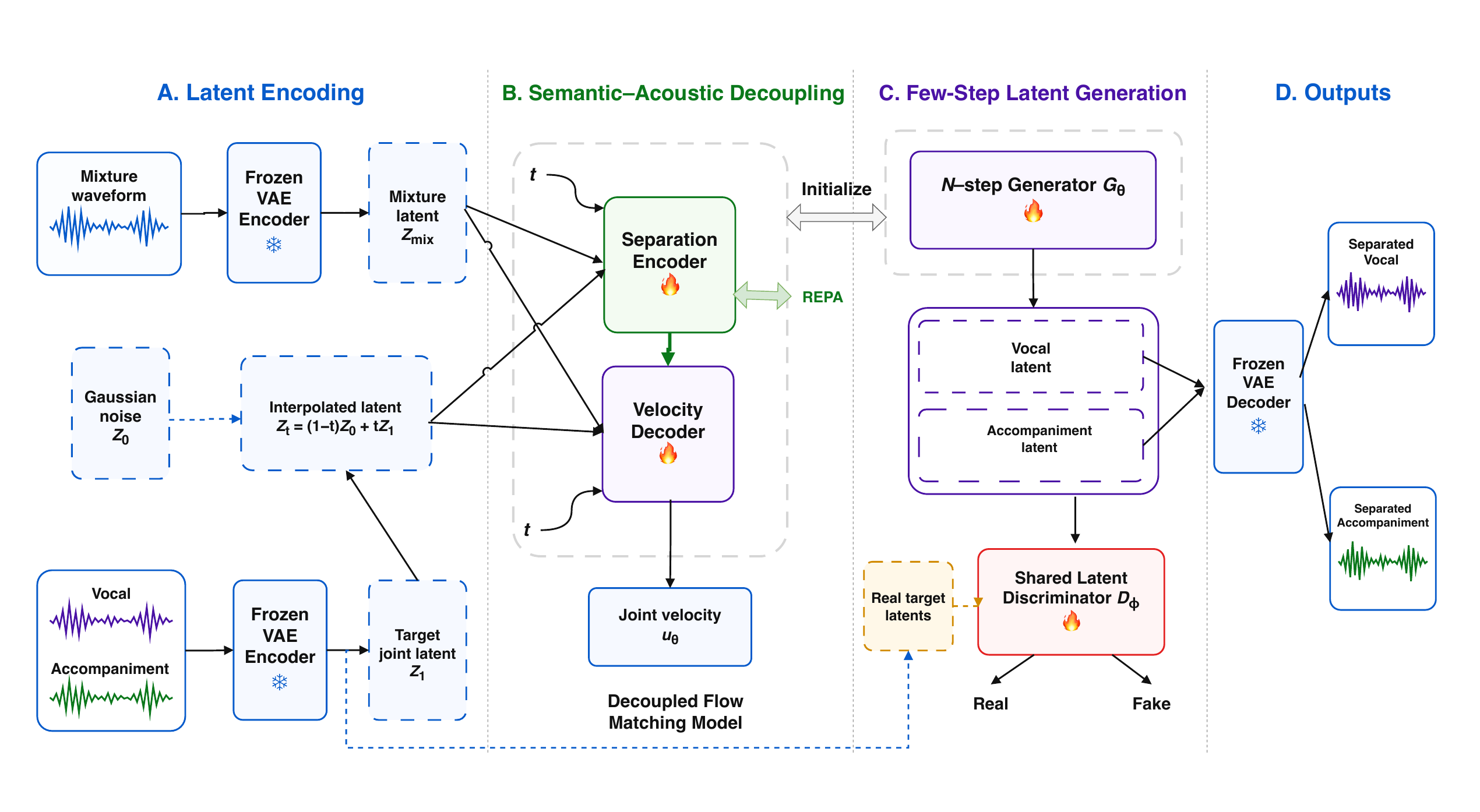}
    \caption{Overview of the proposed framework. (A)~A frozen VAE encodes the mixture and target sources into $\boldsymbol{Z}_{\mathrm{mix}}$ and the joint source latent $\boldsymbol{Z}_{1}$. (B)~The Separation Encoder extracts source-specific conditioning $\boldsymbol{c}$ and is aligned, via $\mathcal{L}_{\mathrm{REPA}}$, with frozen MERT embeddings of the isolated vocal stem. The Velocity Decoder predicts the joint vocal-accompaniment velocity $\boldsymbol{u}_{\theta}$ on the interpolated latent $\boldsymbol{Z}_{t}$ under $\mathcal{L}_{\mathrm{FM}}$. (C)~An $N$-step sampler $G_{\theta}^{N}$ generates source latents, which a shared latent discriminator $D_{\phi}$ refines with $\mathcal{L}_{\mathrm{G}}$. (D)~A frozen VAE decoder maps the generated latents to vocal and accompaniment waveforms.}
    \label{fig:framework}
\end{figure*}

Fig.~\ref{fig:framework} summarizes the proposed framework, including latent encoding, joint source-latent generation, semantic-acoustic decoupling, and latent adversarial post-training.

\subsection{Joint Latent Modeling for Vocal-Accompaniment Separation}

Let $\boldsymbol{x}_{\mathrm{mix}} \in \mathbb{R}^{T}$ denote a mixture waveform vector composed of vocal and accompaniment sources,
\begin{equation}
\boldsymbol{x}_{\mathrm{mix}}
=
\boldsymbol{x}_{\mathrm{vocal}}
+
\boldsymbol{x}_{\mathrm{accomp}},
\end{equation}
where $\boldsymbol{x}_{\mathrm{vocal}},\boldsymbol{x}_{\mathrm{accomp}}\in\mathbb{R}^{T}$ denote the vocal and accompaniment waveform vectors, respectively, and $T$ is the number of waveform samples.

We formulate source separation as joint generative modeling in the latent space of the pretrained VAE. Let $\operatorname{Enc}_{\mathrm{VAE}}$ and $\operatorname{Dec}_{\mathrm{VAE}}$ denote the VAE encoder and decoder. The corresponding latent representations are given by
\begin{equation}
\boldsymbol{Z}_{\mathrm{mix}}
=
\operatorname{Enc}_{\mathrm{VAE}}(\boldsymbol{x}_{\mathrm{mix}}),
\qquad
\boldsymbol{Z}_{s}
=
\operatorname{Enc}_{\mathrm{VAE}}(\boldsymbol{x}_{s}),
\end{equation}
where $s\in\{\mathrm{vocal},\mathrm{accomp}\}$, $\boldsymbol{Z}_{\mathrm{mix}},\boldsymbol{Z}_{s}\in\mathbb{R}^{C\times L}$, and $C$ and $L$ denote the latent channel and temporal dimensions.

We learn a joint flow that models the transformation from an initial Gaussian noise $\boldsymbol{Z}_0$ to the target source latents $\boldsymbol{Z}_1$:
\begin{equation}
\boldsymbol{Z}_0
\sim
\mathcal{N}(\boldsymbol{0},\boldsymbol{I}),
\qquad
\boldsymbol{Z}_1
=
\operatorname{Concat}
\left(
\boldsymbol{Z}_{\mathrm{vocal}},
\boldsymbol{Z}_{\mathrm{accomp}}
\right),
\end{equation}
where $\boldsymbol{Z}_0,\boldsymbol{Z}_1\in\mathbb{R}^{2C\times L}$ and $\boldsymbol{I}$ denotes the identity matrix. Assuming a straight-line transport path, the intermediate state at time $t$ is
\begin{equation}
\boldsymbol{Z}_t
=
(1-t)\boldsymbol{Z}_0
+
t\boldsymbol{Z}_1,
\qquad
t\in[0,1].
\end{equation}

The model is explicitly conditioned on the mixture latent. Its velocity field is written as $\boldsymbol{u}_{\theta}(\boldsymbol{Z}_{t},t;\boldsymbol{Z}_{\mathrm{mix}})$, and the flow-matching objective is
\begin{equation}
\mathcal{L}_{\mathrm{FM}}
=
\mathbb{E}_{(\boldsymbol{Z}_{\mathrm{mix}},\boldsymbol{Z}_{1}),\boldsymbol{Z}_{0},t}
\left[
\left\|
\boldsymbol{u}_{\theta}(\boldsymbol{Z}_{t},t;\boldsymbol{Z}_{\mathrm{mix}})
-
(\boldsymbol{Z}_{1}-\boldsymbol{Z}_{0})
\right\|_{2}^{2}
\right].
\end{equation}
During inference, numerical integration starts at $\boldsymbol{Z}_{0}$ and is conditioned on $\boldsymbol{Z}_{\mathrm{mix}}$ to obtain the generated source latents, which are subsequently decoded into waveform vectors using $\operatorname{Dec}_{\mathrm{VAE}}$.

\subsection{Semantic-Acoustic Decoupling}

Inspired by the functional decoupling in Decoupled Diffusion Transformer (DDT) \cite{wang2026ddt}, we decompose the Diffusion Transformer (DiT)-based flow matching model into two modules for latent music source generation: a Separation Encoder and a Velocity Decoder. The Separation Encoder extracts source-specific representations from the mixture and provides the conditioning information for joint source generation, while the Velocity Decoder predicts the acoustic velocity field conditioned on these representations. This design keeps vocal and accompaniment generation within the same latent flow, but assigns source representation extraction and acoustic velocity prediction to separate modules.

\paragraph{Semantic Representation Alignment.}
The functional decoupled structure above assigns the Separation Encoder the role of learning source-specific representations. To further encourage these representations to capture vocal-relevant semantic information, we employ Representation Alignment for Generation (REPA) \cite{yu2025repa}. Specifically, a frozen pretrained Music undERstanding model with large-scale self-supervised training (MERT) \cite{li2024mert} extracts frame-level embeddings from the isolated vocal stem $\boldsymbol{x}_{\mathrm{vocal}}$, and we align the hidden representations of the Separation Encoder with these vocal MERT embeddings. We use the embedding from the 12th MERT layer as the teacher representation.

Let $\boldsymbol{H}$ denote the hidden representation produced by the Separation Encoder and $\boldsymbol{H}_{\mathrm{MERT}}^{\mathrm{vocal}}\in\mathbb{R}^{B\times M\times C_{\mathrm{MERT}}}$ denote the corresponding frame-aligned vocal MERT embedding, where $B$, $M$, and $C_{\mathrm{MERT}}$ are the batch size, frame length, and MERT feature dimension, respectively. We first map $\boldsymbol{H}$ to the MERT feature dimension using a projection head $g_{\psi}$ with parameters $\psi$, yielding $\tilde{\boldsymbol{H}}=g_{\psi}(\boldsymbol{H})\in\mathbb{R}^{B\times M\times C_{\mathrm{MERT}}}$. We then normalize both representations along the feature dimension and define the alignment loss as
\begin{equation}
\mathcal{L}_{\mathrm{REPA}}
=
1-
\frac{1}{BM}
\sum_{b=1}^{B}
\sum_{m=1}^{M}
\frac{
\tilde{\boldsymbol{H}}_{b,m}^{\top}
(\boldsymbol{H}_{\mathrm{MERT}}^{\mathrm{vocal}})_{b,m}
}{
\|\tilde{\boldsymbol{H}}_{b,m}\|_2
\|(\boldsymbol{H}_{\mathrm{MERT}}^{\mathrm{vocal}})_{b,m}\|_2
}.
\end{equation}

The final training objective combines the flow matching and representation alignment objectives:
\begin{equation}
\mathcal{L}
=
\mathcal{L}_{\mathrm{FM}}
+
\lambda_{\mathrm{REPA}}\mathcal{L}_{\mathrm{REPA}},
\end{equation}
where $\lambda_{\mathrm{REPA}}$ controls the contribution of the representation alignment objective.

\subsection{Adversarial Post-Training in the Latent Space}
\label{sec:latent_adv}

To improve few-step generation, we further fine-tune the flow matching generator with an adversarial objective. We adopt a Flow2GAN-style post-training strategy because it can be applied to an already trained conditional flow separator without redesigning the underlying flow-matching target. In contrast, MeanFlow-style methods modify the training objective to learn average-velocity fields for one-step generation \cite{geng2025meanflow,geng2026improvedmeanflow}, while Distribution Matching Distillation (DMD) is typically formulated as teacher-student distillation from a pretrained diffusion model \cite{yin2024dmd}. The key idea of Flow2GAN \cite{yao2025flow2gan} is to use the discriminator to recover fine-grained and high-frequency details that can be difficult to capture with few-step flow sampling, thereby improving perceptual fidelity while retaining the efficiency of low-step generation. Since the discriminator operates directly on compact latent representations, this refinement is expected to reduce the discriminator-side cost compared with waveform-domain adversarial training.

We use a single shared one-dimensional convolutional discriminator $D_{\phi}$ with parameters $\phi$ for both vocal and accompaniment source latents, following the broader success of adversarial audio generation \cite{kong2020hifigan}. The discriminator consists of four successive downsampling convolutional blocks followed by a convolutional scoring head. The same discriminator parameters are shared across the two source types, with source-specific weights used to balance their contributions.

For each source $s\in\{\mathrm{vocal},\mathrm{accomp}\}$, let $\boldsymbol{Z}_{s}$ and $\hat{\boldsymbol{Z}}_{s}$ denote the reference and generated source latents, respectively. The least-squares generative adversarial network (LSGAN) discriminator objective \cite{mao2017lsgan} is
\begin{align}
\mathcal{L}_{\mathrm{D}}^{(s)}
&:=
\mathbb{E}_{\boldsymbol{Z}_{s}}
\left[
\left(D_{\phi}(\boldsymbol{Z}_{s})-1\right)^2
\right]
+
\mathbb{E}_{\hat{\boldsymbol{Z}}_{s}}
\left[
D_{\phi}(\hat{\boldsymbol{Z}}_{s})^2
\right].
\end{align}

The total discriminator loss is
\begin{align}
\mathcal{L}_{\mathrm{D}}
&:=
\sum_{s\in\{\mathrm{vocal},\mathrm{accomp}\}}
w_{\mathrm{D}}^{(s)}
\mathcal{L}_{\mathrm{D}}^{(s)}.
\end{align}
where $w_{\mathrm{D}}^{(s)}$ controls the contribution of each source to discriminator training.
The adversarial loss of the generator is:

\begin{align}
\mathcal{L}_{\mathrm{adv}}^{(s)}
&:=
\mathbb{E}_{\hat{\boldsymbol{Z}}_{s}}
\left[
\left(D_{\phi}(\hat{\boldsymbol{Z}}_{s})-1\right)^2
\right].
\end{align}

The source-weighted adversarial objective is
\begin{align}
\mathcal{L}_{\mathrm{adv}}
&:=
\sum_{s\in\{\mathrm{vocal},\mathrm{accomp}\}}
w_{\mathrm{G}}^{(s)}
\mathcal{L}_{\mathrm{adv}}^{(s)}.
\end{align}
where $w_{\mathrm{G}}^{(s)}$ controls the contribution of each source to generator training.
To preserve the correspondence between generated and reference source latents, we also use a latent reconstruction constraint,
\begin{align}
\mathcal{L}_{\ell_1}^{\mathrm{latent}}
&:=
\sum_{s\in\{\mathrm{vocal},\mathrm{accomp}\}}
w_{\mathrm{G}}^{(s)}
\left\|
\hat{\boldsymbol{Z}}_{s}-\boldsymbol{Z}_{s}
\right\|_1.
\end{align}

The overall generator objective is
\begin{align}
\mathcal{L}_{\mathrm{G}}
&:=
\lambda_{\mathrm{adv}}\mathcal{L}_{\mathrm{adv}}
+
\lambda_{\ell_1}
\mathcal{L}_{\ell_1}^{\mathrm{latent}}.
\end{align}
where $\lambda_{\mathrm{adv}}$ and $\lambda_{\ell_1}$ control the adversarial and latent reconstruction terms, respectively.

For each target sampling step $N$, we train an $N$-step generator $G_{\theta}^{N}$. For $N>1$, all sampling steps are included in the forward computation and optimized end-to-end through backpropagation, allowing both intermediate states and the final output to be jointly refined. This enables each $G_{\theta}^{N}$ to specifically optimize the trade-off between generation quality and sampling efficiency at its designated number of steps.
\section{Experimental Setup}
\paragraph{Data.}
Each mixture is represented as the sum of two target streams, vocal and accompaniment. 
Following common vocal-accompaniment separation practice, backing vocals are treated as part of the accompaniment in the training data. 
For evaluation, we hold out 50 non-overlapping 20-second segments from the training set and use them exclusively for measuring separation performance.

\paragraph{Evaluation Metrics.}
Because generative separation outputs may be perceptually plausible without being sample-aligned to the reference, we evaluate with phase-insensitive metrics in addition to conventional separation scores. Recent work on generative singing voice separation reports that the Virtual Speech Quality Objective Listener (ViSQOL) and multi-resolution short-time Fourier transform (MR-STFT) loss are suitable for generative separation and coding tasks because they operate on spectrogram features \cite{bereuter2025reliablemetrics}. We therefore use ViSQOL \cite{hines2012visqol,chinen2020visqol} as the primary perceptual metric and MR-STFT loss \cite{yamamoto2020parallelwavegan} as a spectral fidelity metric. Signal-to-distortion ratio (SDR), commonly used in source-separation evaluation campaigns \cite{stoter2018sisec}, is also reported as a conventional separation reference. Higher ViSQOL and SDR indicate better quality, while lower MR-STFT loss indicates better spectral reconstruction.

\paragraph{Training Setup.}
The model is trained in multiple stages. In the initial pretraining stage, the full flow matching model is trained for 135K steps. The model is then further trained for 40K steps under different freezing settings, including continued joint training of both encoder and decoder modules and a schedule that freezes the Separation Encoder while continuing to optimize the Velocity Decoder. For these flow matching stages, we use the AdamW optimizer with a learning rate of $1\times10^{-4}$ and a weight decay of $1\times10^{-3}$.

For few-step generation stage, we further perform Flow2GAN-style adversarial post-training on the same pretrained decoupled latent flow matching model. In the proposed latent-domain setting, the Separation Encoder remains frozen and the Velocity Decoder is optimized for 40K steps using the objective in Section~\ref{sec:latent_adv}. The generator and discriminator learning rates are set to $1\times10^{-5}$ and $4\times10^{-5}$, respectively. We set the source weights $w_{\mathrm{G}}^{(s)}$ and $w_{\mathrm{D}}^{(s)}$ to 1 for both sources, and use unit coefficients for both the adversarial loss and the latent $\ell_1$ loss, i.e., $\lambda_{\mathrm{adv}}=1$ and $\lambda_{\ell_1}=1$. As a comparison, we also apply a waveform-domain Flow2GAN baseline to the same generator: source latents are first decoded by the VAE, and adversarial training then uses the original multi-resolution and multi-period waveform discriminators. For all settings, the pretrained VAE decoder remains frozen throughout training.
\section{Results}

\begin{table*}[!t]
\centering
\caption{Comparison of VAE reconstruction, 20-step flow matching, and 1-/4-step latent Flow2GAN.}
\label{tab:main_results}
\setlength{\tabcolsep}{5.5pt}
\begin{tabular}{clc ccc ccc}
\toprule
& & & \multicolumn{3}{c}{Vocal} & \multicolumn{3}{c}{Accompaniment}\\
\cmidrule(lr){4-6}\cmidrule(lr){7-9}
Row & Method & Steps & ViSQOL & MR-STFT & SDR (dB) & ViSQOL & MR-STFT & SDR (dB)\\
\midrule
1 & VAE reconstruction & -- & 4.201 & 0.846 & 8.792 & 4.168 & 1.069 & 9.520\\
2 & Flow matching & 20 & 3.717 & 1.119 & 4.693 & 3.692 & 1.297 & 7.296\\
3 & Latent Flow2GAN & 1 & 3.826 & 1.100 & 5.895 & 3.676 & 1.298 & 7.318\\
4 & Latent Flow2GAN & 4 & 3.852 & 1.080 & 5.844 & 3.749 & 1.299 & 7.272\\
\bottomrule
\end{tabular}
\end{table*}

\begin{table*}[!t]
\centering
\caption{Waveform- and latent-domain Flow2GAN post-training of the same decoupled flow model under one-step generation.}
\label{tab:latent_vs_waveform}
\setlength{\tabcolsep}{6pt}
\begin{tabular}{clc ccc ccc}
\toprule
& & & \multicolumn{3}{c}{Vocal} & \multicolumn{3}{c}{Accompaniment}\\
\cmidrule(lr){4-6}\cmidrule(lr){7-9}
Row & Method & Steps & ViSQOL & MR-STFT & SDR (dB) & ViSQOL & MR-STFT & SDR (dB)\\
\midrule
1 & Waveform Flow2GAN~\cite{yao2025flow2gan} & 1 & 3.783 & 1.157 & 4.610 & 3.438 & 1.385 & 6.205\\
2 & Latent Flow2GAN & 1 & 3.826 & 1.100 & 5.895 & 3.676 & 1.298 & 7.318\\
\bottomrule
\end{tabular}
\end{table*}

\begin{table*}[!t]
\centering
\caption{Freeze settings after 135K-step flow matching pretraining. Row~1 is the jointly trained baseline. Rows~2 and~3 continue training for 40K steps with different freeze settings.}
\label{tab:freeze_settings}
\setlength{\tabcolsep}{3.7pt}
\begin{tabular}{clccc ccc ccc}
\toprule
& & & & & \multicolumn{3}{c}{Vocal} & \multicolumn{3}{c}{Accompaniment}\\
\cmidrule(lr){6-8}\cmidrule(lr){9-11}
Row & Method & \makecell{Separation\\ Encoder frozen} & \makecell{Velocity\\ Decoder frozen} & Iterations & ViSQOL & MR-STFT & SDR & ViSQOL & MR-STFT & SDR\\
\midrule
1 & Flow matching & No & No & 135K & 3.693 & 1.121 & 4.169 & 3.704 & 1.300 & 6.951\\
2 & Flow matching & No & No & 135K+40K & 3.656 & 1.146 & 4.323 & 3.563 & 1.630 & $-4.467$\\
3 & Flow matching & Yes & No & 135K+40K & 3.717 & 1.119 & 4.693 & 3.692 & 1.297 & 7.296\\
\bottomrule
\end{tabular}
\end{table*}

\paragraph{Main comparison.}
Table~\ref{tab:main_results} reports the main comparison and reveals a clear quality--efficiency trade-off in latent flow matching. Row 1 provides the VAE reconstruction, which corresponds to the decoded ground-truth signal and serves as an upper reference for latent flow matching. Compared with the 20-step Flow Matching baseline in Row 2, the one-step Latent Flow2GAN result in Row 3 improves vocal ViSQOL from 3.717 to 3.826 and vocal SDR from 4.693 dB to 5.895 dB while reducing the sampling budget to a single step. However, accompaniment ViSQOL slightly decreases from 3.692 to 3.676, indicating that few-step adversarial refinement can improve one source more strongly than the other. Row 4 further improves ViSQOL for both sources with four sampling steps, but the SDR gain is not consistent. These results demonstrate that the proposed latent GAN training can substantially reduce the number of inference steps while improving vocal stem quality, without significantly degrading accompaniment stem quality.

\paragraph{Latent versus waveform-domain refinement.}
Table~\ref{tab:latent_vs_waveform} compares two Flow2GAN-style post-training variants of the same decoupled latent flow matching model under a single-step sampling budget. Both start from the pretrained latent generator and differ only in the adversarial domain. In Row 1, waveform-domain Flow2GAN applies adversarial training to VAE-decoded waveforms and uses the multi-resolution (MR) and multi-period (MP) discriminators from the original Flow2GAN design \cite{yao2025flow2gan}. In Row 2, the proposed latent-domain variant keeps adversarial refinement in the VAE latent space and uses a single shared latent discriminator. Despite this simpler discriminator, the latent-domain result consistently achieves better reconstruction quality for both vocal and accompaniment across ViSQOL, MR-STFT loss, and SDR. This suggests that few-step adversarial refinement is more effective in the compact latent space than on decoded waveforms. Nevertheless, a noticeable quality gap remains between the generated outputs and the VAE reconstruction reference, highlighting the difficulty of jointly modeling different musical sources with a shared latent flow.

\paragraph{Decoupled training strategy.}
The decoupled architecture makes it possible to apply a two-stage schedule after the initial flow matching pretraining. Table~\ref{tab:freeze_settings} evaluates this second-stage schedule. Row 1 is the pretrained model after 135K steps of joint training and thus serves as the initial pretraining reference. Row 2 continues to train both the Separation Encoder and the Velocity Decoder for another 40K steps and does not raise this ceiling: ViSQOL and MR-STFT loss degrade for both sources, and accompaniment SDR drops substantially. Row 3 freezes the Separation Encoder while continuing to train the Velocity Decoder, improving vocal ViSQOL from 3.693 to 3.717, vocal SDR from 4.169~dB to 4.693~dB, and accompaniment SDR from 6.951~dB to 7.296~dB. This indicates that prolonged joint updates can interfere with the learned source representation, whereas freezing the Separation Encoder and refining the Velocity Decoder further improves the pretrained flow matching baseline.

\section{Conclusion}
In this work, we formulate vocal-accompaniment separation as joint latent flow matching and decouple source representation extraction from acoustic velocity prediction.
After flow matching pretraining, the latent Flow2GAN-style post-training is conducted for few-step generation. 
In the pretraining stage, we showed that the proposed two-stage pretraining on the decoupled architecture can raise the pretrained flow matching ceiling. 
Also, experiments show that latent adversarial post-training improves model performance under a reduced sampling steps.

\bibliographystyle{unsrt}
\bibliography{main}
\end{document}